\documentclass[aps,prl,showpacs,twocolumn,floatfix,superscriptaddress]{revtex4}
\usepackage{graphicx}
\usepackage{amsmath}
\usepackage{xcolor}

\begin{document}

\title{Optimal noise maximizes collective motion in heterogeneous media}

\date{\today}

\author{Oleksandr Chepizhko}
\affiliation{Department for Theoretical Physics, Odessa National University, Dvoryanskaya 2, 65026 Odessa, Ukraine}
\affiliation{Max Planck Institute for the Physics of Complex Systems, N\"othnitzer Str. 38, 01187 Dresden, Germany}
\affiliation{Laboratoire J.A. Dieudonn{\'e}, Universit{\'e} de Nice Sophia Antipolis, UMR 7351  CNRS , Parc Valrose, F-06108 Nice Cedex 02, France}

\author{Eduardo G. Altmann}
\affiliation{Max Planck Institute for the Physics of Complex Systems, N\"othnitzer Str. 38, 01187 Dresden, Germany}

\author{Fernando Peruani}\email{Peruani@unice.fr}
 \affiliation{Laboratoire J.A. Dieudonn{\'e}, Universit{\'e} de Nice Sophia Antipolis, UMR 7351  CNRS , Parc Valrose, F-06108 Nice Cedex 02, France}

\begin{abstract}
We study the effect of spatial heterogeneity on the collective motion of self-propelled particles (SPPs). 
The heterogeneity is modeled as a random distribution of either static or diffusive obstacles, which the SPPs avoid while trying to align their movements. 
We find that such obstacles have a dramatic effect on the collective dynamics of usual SPP models. 
In particular, we report about the existence of an optimal (angular) noise amplitude that maximizes  collective motion. 
%
%
We also  show that while at low obstacle densities the system exhibits long-range order, in strongly heterogeneous media collective motion is quasi-long-range and exists only for  noise values in between two critical noise values, with the system being disordered at both, large and low noise amplitudes. 
Since most real system have spatial heterogeneities, the finding of an optimal noise intensity has immediate practical and fundamental implications for the design and evolution of collective motion strategies. 
\end{abstract}


\maketitle

Most examples of natural systems, if not all,  where collective motion occurs in the wild,  take place in heterogeneous media.    
Examples can be found at all scales. Microtubules driven by molecular motors form complex patterns inside the cell where the space is filled by organelles and vesicles~\cite{alberts}. 
Bacteria exhibit complex collective behaviors, e.g. swarming, in heterogeneous environments such as the soil or highly complex tissues such as in the gastrointestinal tract~\cite{dworkin}.   
At a larger scale,  herds of mammals migrate long distances traversing rivers, forests, etc~\cite{holdo2011}.
Despite of these evident facts, little is known at both levels, experimental as well as theoretical, 
 about the impact that an heterogeneous medium may have on the self-organized collective motion~\cite{vicsek2012}.  
For instance, most collective motion  experiments have been performed on homogeneous arenas~\cite{vicsek2012}, from microtubules moving on fixed carpet of molecular motors~\cite{schaller2010}, 
bacteria swarming on surfaces~\cite{zhang2010, peruani2012}, to marching locusts~\cite{romanczuk2009}, and including fabricated self-propelled systems~\cite{kudrolli2008, deseigne2010}. 
Not surprisingly, most theoretical efforts have also focused on homogeneous media~\cite{vicsek2012, marchetti2012c},   
from the pioneering work of Vicsek et al.~\cite{vicsek1995} to the detailed study of symmetries and large-scale patterns in self-propelled particle systems~\cite{gregoire2004,  ginelli2010, peruani2008, peruani2011, marchetti2012b, baskaran2012, TonerTuPRL1995, TonerTuPRE1998, RamaswamyEuroPhLett2003}, 
where the transition to collective motion is reduced  to the competition between a local aligning interaction
and a noise. 

\begin{figure}[b]
\centering
\resizebox{\columnwidth}{!}{\rotatebox{0}{\includegraphics{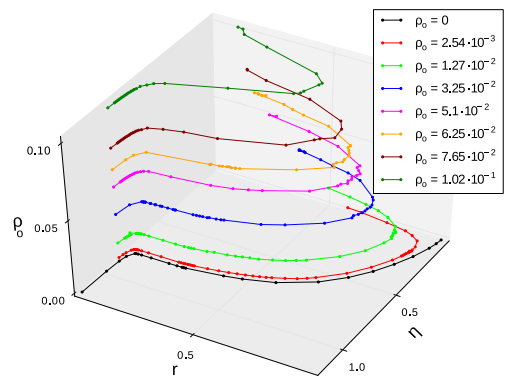}}}
\caption{(color online). Optimal noise amplitude. 
Order parameter $r$ as a function of noise strength $\eta$ and obstacle density $\rho_o$. 
%
Data corresponding to $L = 140$, $D_o=0$, and $\rho_b=1$.} \label{fig:optimalNoise}
\end{figure}

Here, we show through a simple model that the presence of even few either static or diffusive heterogeneities changes qualitatively the collective motion dynamics. 
In particular, we find that there is an optimal noise amplitude that maximizes collective motion, while in an homogeneous medium such an optimal does not exist, see Fig.~\ref{fig:optimalNoise}. 
For weakly heterogeneous media ( i.e., low obstacle densities)  we observe that the transition to collective motion exhibits a unique critical point below, which the system exhibits long-range order, 
as in  homogeneous media.  
For strongly heterogeneous media (high obstacle densities), we find on the contrary that there are two critical points, with the system being disordered at both, large and low noise amplitudes, 
and exhibiting only quasi-long-range order in between these critical points.  
The finding of an optimal noise that maximizes  self-organized collective motion may 
help to understand and design migration and navigation strategies in either static or fluctuating heterogeneous media, which in turn may 
shed some light on the adaptation and evolution of stochastic components in natural systems that exhibit collective motion, for instance, concerning the bacterial tumbling rate. 

\begin{figure*}
\centering
\resizebox{15.5cm}{!}{\rotatebox{0}{\includegraphics{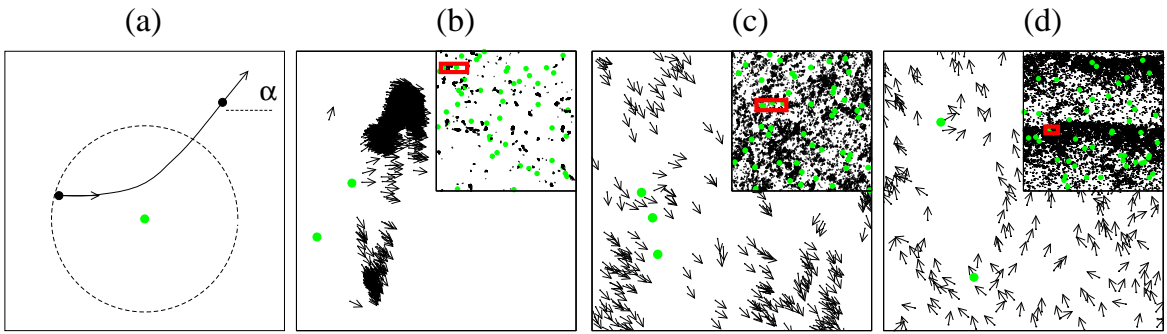}}}
\caption{(color online). (a) Details of the interaction between a SPP and an obstacle ($\eta = 0.1$). The dashed circle represents the interaction area, of radius $R_o$,  the solid (black) curve corresponds to  the particle trajectory, and $\alpha$ is the scattered angle.  
(b), (c) and (d) illustrate the different phases exhibited by the system with $D_o=0$ and $\rho_o = 2.55\,\cdot10^{-3}$ at the microscopic and macroscopic level: 
(b) clustered phase, $\eta = 0.01$ with order parameter $r=0.58$, (c) homogeneous (ordered) phase, $\eta =0.3$ with $r=0.97$, and (d), band phase, $\eta = 0.6$ with $r=0.73$.
Insets correspond to snapshots of the entire system, where the red box inside them indicates the system area that is shown on main panel. For movies illustrating these phases see~\cite{movie}.
} \label{fig:fourPhases}
\end{figure*}

{\it Model definition.--}
We consider a continuum time model for $N_b$ SPPs moving in a two-dimensional space, with periodic boundary conditions, of linear size $L$. 
%
SPPs interact among themselves via a (local) ferromagnetic velocity alignment as in~\cite{vicsek1995}.  
Spatial heterogeneity is modeled by the presence of either fixed or diffusive obstacles. 
The new element in the equation of motion of the SPPs is given by 
  the obstacle avoidance interaction by which SPPs turn away from obstacles whenever they are at a distance equal or less than $R_o$ from them.  
The implementation of this rule is analogous to the archetypical (discrete) collision avoidance rule introduced in~\cite{chate2004}. 
In the over-damped limit, we express the equations of motion of the $i$-th 
particle as:
\begin{eqnarray}
\label{eq:evol_x}
\dot{\mathbf{x}}_i &=& v_0 \mathbf{V}(\theta_i) \\
\label{eq:evol_theta}
\dot{\theta}_i       &=&   g(\mathbf{x}_i) \left[ \frac{\gamma_b}{n_b(\mathbf{x}_i)}  \!\!\!\!\!\! \sum_{\quad |\mathbf{x}_i-\mathbf{x}_j|<R_b} \!\!\!\!\!\!\! \sin(\theta_j - \theta_i) \right] + \\
\nonumber &&  + \left[ \frac{ \gamma_o}{n_o(\mathbf{x}_i)}  \!\!\!\!\!\! \sum_{\quad |\mathbf{x}_i-\mathbf{y}_k|<R_o} \!\!\!\!\!\!\! \sin(\alpha_{k,i} - \theta_i) \right] +  \eta  \xi_{i}(t)  \, , 
\end{eqnarray}
where the dot denotes temporal derivative, $\mathbf{x}_i$ corresponds to the position of the $i$-th particle, $\theta_i$ to its moving direction, and $\mathbf{y}_k$ is the position of the $k$-th obstacle. 
In Eq.~(\ref{eq:evol_x}),  $v_{\rm 0}$ is the active particle speed and  $\mathbf{V}(\theta)\equiv (\cos(\theta),\sin(\theta))^T$.   
The interaction SPP-SPP is defined by two parameters,  the angular (relaxation) speed $\gamma_b$ and the interaction radius $R_b$. 
Similarly, the interaction SPP-obstacle is determined by $\gamma_o$ and $R_o$. 
The term $n_b(\mathbf{x}_i)$ ($n_o(\mathbf{x}_i)$) corresponds to the number of SPPs (obstacles) that are located at a distance less or equal than $R_b$ ($R_o$) from $\mathbf{x}_i$. 
In the second sum in  Eq.~(\ref{eq:evol_theta}),  the term $\alpha_{k,i}$ denotes the angle, in polar coordinates, of the vector $\mathbf{x}_i - \mathbf{y}_k$. 
The additive white noise is characterized by an amplitude $\eta$ and obeys  
 $\langle \xi_{i}(t) \rangle = 0$ and $\langle \xi_{i}(t) \xi_{j}(t') \rangle =\delta_{i,j} \delta(t-t')$. 
%
The term $g(\mathbf{x}_i)$ in Eq.~(\ref{eq:evol_theta})  controls the strength of the alignment with respect to obstacle avoidance. 
For instance, $g(\mathbf{x}_i) =  \left[1-\Theta[n_o(\mathbf{x}_i)]\right]$ with $\Theta[n]=1$ if $n>0$, and $0$ otherwise (switching rule), represents 
a scenario in which SPPs stop aligning in the presence of an obstacle, analogous to the hardcore repulsion rule introduced in~\cite{chate2004}. 
We also consider a simpler scenario with  $g(\mathbf{x}_i) = 1$  (no switching rule) where particles never stop aligning to neighbors.  
Finally, obstacles are either fixed in space, or diffuse around with a diffusion coefficient $D_o$. 
For simplicity, we initially fix $R_b=R_o=1$, $\gamma_b=\gamma_o=1$,  $\rho_b= N_b/L^2 =1$,  $v_0=1$, and $D_o=0$  (with a discretization time $\Delta t=0.1$), and use the switching rule.  
Other scenarios are discussed at the end. 

If $\gamma=0$, equations~(\ref{eq:evol_x}) and~(\ref{eq:evol_theta}) define a system of non-interacting persistent random walkers. 
%
For $\gamma>0$ and $N_o=0$, Eq.~(\ref{eq:evol_x}) and~(\ref{eq:evol_theta}) reduce to a continuum time version of the Vicsek model (VM)~\cite{vicsek1995} as proposed in~\cite{peruani2008}. 
%
%
It is for $\gamma>0$ and $N_o\geq 0$ that we observe a completely new behavior, since now  the SPPs not only align among themselves but also avoid 
 obstacles by turning away from them, with a characteristic turning time given by $1/\gamma$.  
%
%
Fig.~\ref{fig:fourPhases} illustrates the new aspects of the  collective behavior, as well as a typical interaction between an obstacles and a SPP. 
%

{\it Optimal noise.--} 
To characterize the macroscopic collective motion we use the following order parameter:  
\begin{eqnarray}\label{eq:orderparam_f}
 r =   \langle r(t) \rangle_t  = \langle  \left| \frac{1}{N_b} \sum_{i=1}^{N_b} e^{i \theta_i(t)} \right| \rangle_t  \,,  
\end{eqnarray}
 where $\langle \hdots \rangle_t$ denotes temporal average.  
%
Fig. \ref{fig:optimalNoise} shows $r$ versus the angular noise $\eta$ for various obstacle densities $\rho_o=N_o/L^2$. 
The curve $\rho_o=0$ corresponds to the continuum time VM and 
as  the noise amplitude $\eta$ is decreased below a critical amplitude $\eta_{c1}$, $r$ monotonically increases, with   $r \to 1$ as $\eta \to 0$~\cite{vicsek2012}. 
Here,  we find that for $\rho_o>0$ the scenario is qualitatively different and $r$  exhibits a non-monotonic behavior with $\eta$. Moreover, we observe that there is an optimal angular noise amplitude  $\eta_M$ at which $r$ reaches a 
maximum value.    
The relevance of this striking result is clear, due to the presence of a random distribution of obstacles, there exists an angular noise $\eta_M$ that maximizes the collective motion.  
Notice that  in a simple model of particles driven in opposite directions it has been reported also the existence of an ``optimal" noise, but in this case,  
contrary to what we report here, it freezes particle motion~\cite{vicsek2000}. 
Fig. \ref{fig:optimalNoise} shows that the system is disordered, without exhibiting collective motion  for  $\eta>\eta_{c1}$. 
Collective motion and orientational order increase as $\eta$ is decreased from  $\eta_{c1}$ to $\eta_{M}$. 
Counterintuitively, decreasing $\eta$ further hinders collective motion. 
If the density of obstacles  $\rho_o$ is large enough, we find that unambiguously the system becomes fully disordered again but this time for $\eta << \eta_{M}$.  
The remarkable fact is that there is a second, nonzero, critical angular noise amplitude $\eta_{c2}$ at large enough densities $\rho_o$. 
%

\begin{figure}
\centering
\resizebox{\columnwidth}{!}{\rotatebox{0}{\includegraphics{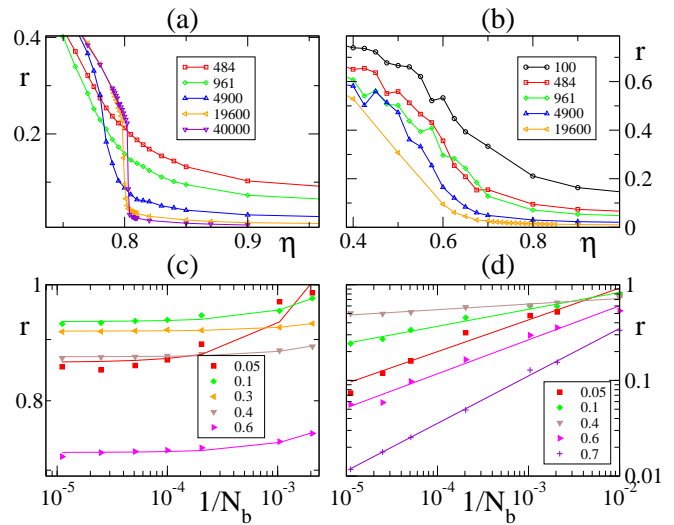}}}
\caption{(color online). Finite size scaling. Order parameter $r$ vs. angular noise $\eta$ for various system sizes $N_b$ (color coded) for $\rho_o = 2.55\,\cdot10^{-3}$ in (a) and  $\rho_o = 0.102$ in (b). 
%
%
The scaling of $r$ with system size $N_b$ at fixed angular noise $\eta$ (color coded) is shown in (c) and (d) for the obstacle densities corresponding to (a) and (b), respectively.  
The solid curves correspond to exponential fittings in (c) and power-laws in (d), which suggests the presence of LRO and QLRO, respectively.  
} \label{fig:FiniteSize}
\end{figure}

{\it Order-disorder transitions.--} 
At low densities $\rho_o$, the obtained numerical data suggests that for $\eta \leq  \eta_{c1}$   the system exhibits 
long-range order (LRO). 
Increasing the system size, while keeping  densities $\rho_b$ and $\rho_o$ constant, we observe that the transition becomes sharper with system size,  Fig. \ref{fig:FiniteSize}(a). 
%
%
%
The transition at $\eta_{c1}$ is accompanied by the emergence of traveling high density structures, i.e., moving bands as observed in the VM~\cite{chate2004}.  
%
%
Bands are observed only close to $\eta_{c1}$ and at 
the optimal angular noise $\eta_{M}$, they have always disappeared.
On the other hand, as the density of obstacles $\rho_o$ is increased,  bands contain less particles, while the background density of SPPs increases, to the point that for large values of $\rho_o$ bands are no  longer  observed. 
%

The existence of LRO implies that  for a fixed $\eta$ value, $r$ should tend to an asymptotic value larger than $0$ as the system size $N_b$ goes to infinity. 
A useful way to estimate this limit is to plot $r$ as function of the inverse system size $y$, with $y=1/N_b$, and extrapolate the behavior of $r$ when $y \to 0$. 
This is shown in Fig. \ref{fig:FiniteSize}(c) for $\rho_o = 2.55\cdot10^{-3}$, where the solid curves correspond to fittings with exponentials, i.e., $r \sim r_{\infty}(\eta) \exp(A(\eta) N_b)$.  
Such a scaling strongly suggests the existence of LRO  for  $\eta < \eta_{c1}$  at low $\rho_o$ densities.  

\begin{figure}
\centering
\resizebox{7.5cm}{!}{\rotatebox{0}{\includegraphics{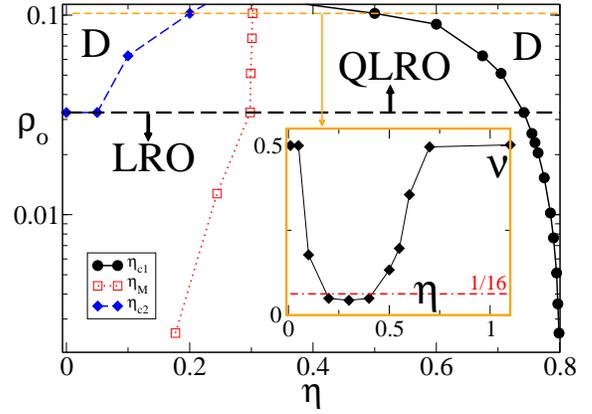}}}
\caption{(color online). Phase diagram. The solid black curve with dots corresponds to the critical noise amplitude $\eta_{c1}$  and sets the boundary between a disordered (D) and an ordered phase. 
The ordered phase, below the horizontal dashed black line, corresponds long-range order (LRO), while below it, to quasi-long-range order (QLRO). 
Above the horizontal dashed black line, there is a second critical point, $\eta_{c2}$, indicated by the blue diamond curve. 
The dotted red curve indicates the position of the optimal noise strength $\eta_{M}$. 
The inset shows the behavior of the finite-size scaling exponent $\nu$ in Eq.~(\ref{eq:scaling_QRL}) with the noise amplitude $\eta$ for $\rho_o=0.102$, which evidences the presence of the two critical points, see text and~\cite{parameters}.} \label{fig:PD}
\end{figure}

At higher densities, the system behavior is remarkably different. 
Fig. \ref{fig:FiniteSize}(b) shows that this time as the system size $N_b$ is increased, the transition becomes smoother, with 
the order parameter $r$ decreasing with system size for all $\eta$ value. 
We find that  $r$ obeys the following scaling with 
system size $N_b$: 
\begin{eqnarray}\label{eq:scaling_QRL}
r \propto N_b^{-\nu(\eta, \rho_o)}   \,,    
\end{eqnarray}
with $\nu(\eta, \rho_o)>0$, Fig. \ref{fig:FiniteSize}(d). 
Though this finding is somehow reminiscent of an equilibrium Kosterlitz-Thouless (KT) transition~\cite{kosterlitz1973}, there are various fundamental differences. 
%
In first place, $\nu$ exhibits a non-monotonic behavior with $\eta$, with a minimum at $\eta_ M$, and $\nu = 1/2$ at low and high $\eta$ values,   see inset in Fig. \ref{fig:PD}. 
Such a scaling corresponds to a fully disordered phase and indicates that in addition to $\eta_{c1}$,  there is a second critical point $\eta_{c2}$ for low $\eta$ values.  
In analogy with the KT transition, we defined $\eta_{c2}$ as the angular noise at which  $\nu = 1/16$. 
When $0 < \nu < 1/16$, we say that the system exhibits quasi-long range-order (QRLO). 
%
%
We stress that  $\nu \to 1/2$ for nonzero $\eta$-values below $\eta_{c2}$, while $\nu$  reaches its minimum value as $\eta \to \eta_{M}$. 
In conclusion, the numerical data for high obstacle densities $\rho_o$ is consistent with QLRO for $\eta_{c2} \leq \eta \leq \eta_{c1}$. 
This means that at some intermediate density $\rho_o^*$, which we roughly estimate around $\rho_o^* = 0.03$, there is a transition from LRO to QLRO. 
%

\begin{figure}
\centering
\resizebox{\columnwidth}{!}{\rotatebox{0}{\includegraphics{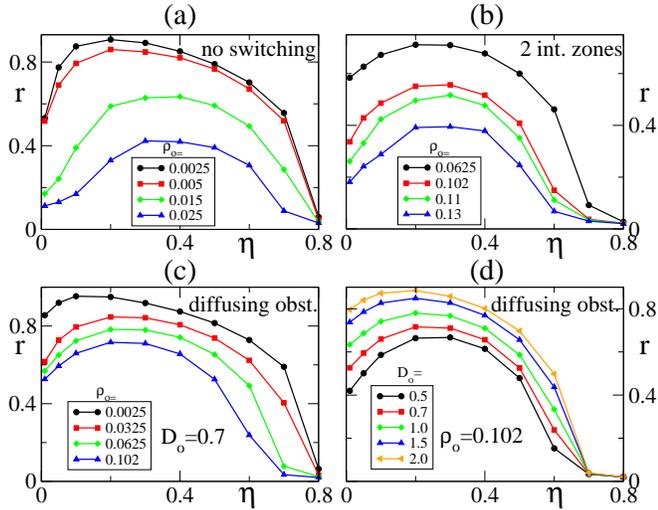}}}
\caption{(color online). 
Robustness and generality of results. 
The same macroscopic behavior is observed in various variations of the model. 
Data obtained with: (a)  the no switching rule, i.e., $g=1$, 
(b)  two interacting zones (with $R_o=0.5$, $\gamma_o=5$ and $R_b=\gamma_b=1$), 
while (c) and (d) correspond to diffusing obstacles, i.e., $D_o>0$. 
When parameters are not specified, they correspond to those used previously~\cite{parameters}. 
  } \label{fig:robustness}
\end{figure}

{\it Phases and physical interpretation.--} 
We have seen that when $\rho_o>0$  the order parameter $r$ exhibits a maximum at $\eta_M$. 
This means that we can find values of $\eta$ to the left and to the right of $\eta_M$ that lead to the same value of the order parameter  $r$.  
The next logical question is whether we can say something regarding the state of the system for two different $\eta$-values that lead to the same value of  $r$.  
To the right of  $\eta_{M}$ and close to  $\eta_{c1}$ particles organize into bands, Fig. \ref{fig:fourPhases}(d). 
To the left of $\eta_{M}$ and close to $\eta_{c2}$, on the other hand, particles form very dense clusters and freely moving particles are rarely observed. 
%
When these dense clusters collide with an obstacle, they often split into two or more fragments that are deflected away, see Fig. \ref{fig:fourPhases}(b) and cluster phase movie in~\cite{movie}. 
The new formed sub-clusters tend to move in uncorrelated directions. 
%
%
%
The dynamics is such that while a cluster recruits particles and other clusters in between collisions, it breaks into very cohesive sub-clusters that move in different direction at each collision with an obstacle, 
with each sub-cluster experiencing a similar fate. 
As result of this process, the SPPs cannot form a highly ordered particle flow. 
But if $\eta$ is increased, clusters are less cohesive and quickly spread out. This fast spreading of clusters  allows sub-clusters to quickly reconnect and orientational order information  
is more efficiently distributed  across the system, see Fig. \ref{fig:fourPhases}(c).  
On the other hand, if we keep on increasing $\eta$,  the noise ends up being  too strong for the alignment strength $\gamma$ and the system becomes disordered again.  
%

{\it Concluding remarks.--}
The same macroscopic behavior is observed in various SPP systems, which provides a strong evidence of the robustness and generality of the reported results. 
In particular, the existence of an optimal noise seems to be rooted in the fact that  
a certain amount of noise facilitates, in the presence of obstacles, the exchange of particles and information among clusters, which in turn promotes  the emergence of large correlations in the system. 
Fig.~\ref{fig:robustness} shows in (a) that the use of the ``no switching" interacting rule between SPP-obstacles, i.e.  $g(\mathbf{x}_i)=1$, results in the same behavior,
in (b) that two interacting zones, for instance, a larger alignment zone with a smaller and faster repulsion zone, mimicking a hardcore repulsion as proposed in~\cite{chate2004}, 
do not alter the obtained results, and in (c) and (d) that the same macroscopic behavior is also observed with diffusing obstacles. 
This last observation is of particular relevance and extends the obtained results to fluctuating environments, which are of particular relevance in biological contexts such as 
the self-organization of microtubules inside the cell~\cite{alberts} or  bacterial self-organization in hostile environments where either poisonous chemicals or bacteria predators as lymphocytes diffuse around~\cite{dworkin}. 
We notice that the stronger the diffusion $D_o$, the weaker the effect, with an increase of $D_o$ playing a similar role as a decrease of $\rho_o$, Fig.~\ref{fig:robustness}(d).

Our analysis reveals -- up to the system sizes we manage to explore -- that the presence of heterogeneous media leads to an unexpectedly complex phase diagram, as summarized in Fig.~\ref{fig:PD}. 
%
%
The most remarkable finding is the qualitative change of behavior -- in a two dimensional system with continuum symmetry -- from long-range order (LRO)  and a unique critical point  ($\eta_{c1}$), at low $\rho_o$, to quasi-long-range  order (QLRO) and two critical points ($\eta_{c1}$ and $\eta_{c2}$), at high  $\rho_o$. 
Notice that  QLRO occurs with particles and interactions maintaining their polar symmetry and at finite densities, while QLRO in homogeneous SPP systems has been 
found with particles and interactions exhibiting both apolar symmetry~\cite{chate2006}, as well as with metric interactions but in the zero density limit only~\cite{ginelli2010b}. 
%
Finally, there is a qualitative difference to previous ``noise-induced order" examples~\cite{vicsek2000, chakrabarti2003, chakrabarti2004, dzubiella2002, rosato1987, herrmann1992, herrmann1995}:
the increase of order occurs here without requiring an external field or driving (and it is not induced by boundary conditions).  
%
A direct comparison with lane formation in systems with two populations of particles driven by an external field in opposite directions~\cite{vicsek2000, chakrabarti2003, chakrabarti2004, dzubiella2002} reveals further important  differences~\cite{remark2lanes}, with the density of opposite moving particles playing the role of our noise and the strength of the external field as the inverse of our density of obstacles (cf.~\cite{chakrabarti2004}).  

In summary, we have reported about:
 1) the existence of an optimal noise for self-organized collective motion in heterogeneous media, 
 2) a transition from LRO to QLRO in 2D, 
 3) QLRO in SPP systems at finite density with  particles and interactions exhibiting polar symmetry, 
and 4) an example of noise-induced order without requiring an external field.





Numerical simulations have been performed at the `Mesocentre SIGAMM' machine, hosted by Observatoire de la C{\^o}te d'Azur.

\bibliographystyle{apsrev}

\begin{thebibliography}{23}
\bibitem{remark2lanes} The comparison requires a suitable Galilean transformation. Notice that the external field breaks the symmetry.
\end{thebibliography}

\end{document}